\documentstyle[12pt]{article}
\newcommand{\spone}{0.9}

\newcommand{\sptwo}{1.4}
\newcommand{\spthree}{2.4}

\newcommand{\singlespace}{\edef\baselinestretch{\spone}\Large\normalsize}

\newcommand{\doublespace}{\edef\baselinestretch{\sptwo}\Large\normalsize}
\newcommand{\threespace}{\edef\baselinestretch{\spthree}\Large\normalsize}
\begin{document}
\doublespace
\threespace
\singlespace
\begin{center}
{\bf Ground-State of  Charged Bosons Confined in a Harmonic Trap}\\
\renewcommand\thefootnote{\fnsymbol{footnote}}
{Yeong E. Kim \footnote{ e-mail:yekim$@$physics.purdue.edu} and
Alexander L. Zubarev\footnote{ e-mail: zubareva$@$physics.purdue.edu}}\\
Department of Physics, Purdue University\\
West Lafayette, Indiana  47907\\
\end{center}
\begin{quote}
We study a system composed of $N$ identical charged bosons confined in a 
harmonic
trap. Upper and lower energy bounds are given. It is shown in the large N limit
that
the ground-state energy is 
determined within an accuracy of $\pm 8\%$ and that the mean field theory
provides 
a reasonable result with relative error of less than 16\% for the binding energy.
\end{quote}

\vspace{5mm}

\vspace{55 mm}

\pagebreak

{\bf I. Introduction}
\vspace{8pt}

We study a system composed of N identical bosons interacting via the Coulomb 
repulsive force, which are confined in an isotropic harmonic trap.

Investigations of
charged Bose gases have been reported in number of papers [1-7].
In recent papers [6, 7],
the mean field theory for bosons in the form given in Ref.[8] was used
to describe the ground state of a bosonic Thomson atom.
Equivalence of the Coulomb systems in a harmonic
trap to the Thomson atom model (``raisin cake" model) [9] was discussed in
Refs.[6,10,11]. The model approximately simulates a number of physical situations  such as
systems of ions in a three-dimensional trap (radio-frequency or
Penning trap) [10,11], electrons in
quantum dots [12, 13], etc.

Since  no exact general solution of the $N$-body problem has been found, 
to investigate validity of the mean-field approximation for the case of systems of charged bosons confined in a trap, we propose in this paper to compare the mean-field
energy with lower and upper bounds. Such approach was used to establish the
asymptotic accuracy  
 of the Ginzburg-Pitaevskii-Gross ground state energy
for dilute neutral Bose gas with repulsive interaction [14].

We find that our lower and upper bounds provide the actual value of 
ground-state 
 energy within $\pm 8\%$ accuracy. We also show that, for the case of
 large $N$, 
the mean-field theory is a reasonable approximation with  
a relative error of less then 16 \% for the binding energy.

The paper is organized as follows. In Section II, we describe an outline of 
the mean-field method. Energy and single-particle density are found 
analytically in the large N limit. In Section III, we generalize 
a lower-bound method developed by Post and Hall [15] for the case 
of charged bosons confined in a
harmonic trap. In Section IV, we describe the strong coupling pertubative
expansion method. In
Section V, we describe our calculation of upper bounds using the effective linear 
two-body equation (ELTBE) method [16]. In Section VI we consider the Wigner
crystallization regime.
A summary and conclusions are given in Section VII.
\vspace{18pt}

{\bf II. Mean-Field Method}
\vspace{8pt}

To describe ground-state properties of a system of interacting bosons
confined in a harmonic trap, we start from the mean-field theory for bosons
in the following  form  given in Ref.[8]
$$
[-\frac{\hbar ^2}{2m}\bigtriangleup+V_t(\vec{r})+(N-1)V_H(\vec{r})]\Psi(\vec{r})
=\mu \Psi(\vec{r}),
\eqno{(1)}
$$
where $\Psi(\vec{r})$ is the normalized ground-state wave function,
$V_t(\vec{r})=m \omega^2 r^2/2$ is a harmonic trap potential
 with $r^2= x^2+
y^2+z^2$,
$V_H(\vec{r})=\int d\vec{r}^{\prime}V_{int}(\vec{r}-\vec{r}^{\prime})$ $~$ $
\mid \Psi(\vec{r}^{\prime})\mid^2$ is the Hartree potential with
an interacting potential
$V_{int}(\vec{r})$, and $N$ is  number of particles in a trap. The
chemical potential $\mu$ is
related to the mean-field ground-state energy $E_M$ and particle number $N$ by the general
thermodynamic identity
$$
\mu=\frac{\partial E_M}{\partial N},
\eqno{(2)}
$$
for $N \rightarrow \infty$,
where the mean-field ground-state energy $E_M$ is given by
$$
E_M=N[<\Psi \mid-\frac{\hbar ^2}{2m}\bigtriangleup \mid \Psi >+
<\Psi \mid V_t \mid \Psi> + \frac{N-1}{2}<\Psi \mid V_H \mid \Psi >].
\eqno{(3)}
$$
We note that the mean-field theory, Eq. (1), can not describe the Wigner
crystalization regime [17] (see also Ref. [6]).

We introduce dimensionless units by making the following transformations:
(i) $ \vec{r} \rightarrow a \vec{r}$, where
$a=\sqrt{\hbar/(m \omega)}$, and  (ii) the energy and chemical
potential
 are measured in units of $\hbar \omega$.

Using the  above dimensionless notation, we can rewrite Eq. (1) as
$$
[-\frac{1}{2}\bigtriangleup+ \frac{r^2}{2}+
(N-1) \int d \vec{r^{\prime}}V_{int}(\vec{r}-\vec{r^{\prime}})
\mid \Psi(\vec{r^{\prime}})\mid^2]\Psi(\vec{r})=\mu \Psi(\vec{r}).
\eqno{(4)}
$$

In the limit $N\gg$ 1, the nonlinear Schr\"odinger equation (4) can be
 simplified
by omitting the kinetic energy, yielding the following integral equation
$$
\frac{r^2}{2}+N \int d \vec{r^{\prime}}V_{int}(\vec{r}-\vec{r^{\prime}})
\mid \Psi(\vec{r^{\prime}})\mid^2=\mu,
\eqno{(5)}
$$
where $r^2<2\tilde{\mu}$,
and $\mid \Psi(\vec{r})\mid^2=0$, if
$r^2>2\tilde{\mu}$,
$\tilde{\mu}$
is to be determined
from the minimum of the energy functional
$$
E_M=\frac{N}{2} \int \mid \Psi(\vec{r})\mid^2 r^2 d \vec{r}+
\frac{N^2}{2}\int
\mid \Psi(\vec{r})\mid^2 \mid \Psi(\vec{r^{\prime}})\mid^2
V_{int}(\vec{r}-\vec{r^{\prime}}) d \vec{r} d \vec{r^{\prime}}.
$$
 This method (Eq. (5)) is another possible  implementation of the
Thomas-Fermi treatment of neutral, dilute vapors [18,19].
For review of the Thomas-Fermi theory of atoms see Ref.[20].

To make a proper choice for the large-$N$ limit of the Hamiltonian
for bosons
interacting via the
Coulomb potential
$$
V_{int}(r)=\frac{\gamma_c}{r},
\eqno{(6)}
$$
with $\gamma_c=Z^2\alpha \sqrt{mc^2/(\hbar \omega)}>0$,
we rescale variables $\vec{r}=(N \gamma_c)^{1/3} \vec{z}$.
Now we can rewrite Eq. (4) as
$$
[-\frac{\epsilon}{2}\bigtriangleup+
\frac{z^2-R^2}{2}+
\int \frac{d \vec{z^{\prime}}}{\mid\vec{z}-\vec{z^{\prime}}\mid}
\mid \Psi(\vec{z^{\prime}})\mid^2]\Psi(\vec{z})=0,
\eqno{(7)}
$$
where $R^2=\frac{2 \mu}{(N \gamma_c)^{2/3}},$  $\epsilon=\frac{1}{(N \gamma_c)^{
4/3}}$, and $N\gg$ 1.

In the case  $N \gamma_c\gg$ 1,
the solution of Eq. (5) is found to be
$$
\mid \Psi(\vec{r})\mid^2=\frac{3}{4\pi N \gamma_c}
\theta(2 \tilde{\mu}-r^2),
\eqno{(8)}
$$
where $\theta$ denotes the unit positive step function, and
$$
\tilde{\mu}= \frac{\mu}{3}.
\eqno{(9)}
$$

Straightforward calculations with $\mid \Psi(\vec{r})\mid^2$ from Eq. (8) yield
$$
\mu=\frac{3}{2}(\gamma_c N)^{2/3},
$$
$$
E_M=\frac{9}{10}(\gamma_c)^{2/3} N^{5/3}.
\eqno{(10)}
$$

Eq. (8) is obtained by neglecting $\frac{\epsilon}{2}\bigtriangleup \Psi$
term in Eq. (7) and  provides an accurate description of  the exact solution 
where
 the  gradients of   the wave function are small. In a  boundary layer of  a
narrow region near surface, the approximation (8) breaks down. We expect that
 the thickness of this  boundary layer approaches zero as $\epsilon \rightarrow
0$.
Recent
numerical calculations [6] support our analytical results.
Eq. (10) 
provides an upper bound for the ground state energy in the large $N$ limit ($ N\gg1$, and $N\gamma_c \gg1$).
\vspace{18pt}

{\bf III. Lower Bounds}
\vspace{8pt}

In this section, we consider $N$ identical charged bosons
confined in a
harmonic isotropic trap with the following Hamiltonian

$$
H=-\frac{1}{2}\sum_{i=1}^{N} \Delta_{i}+\frac{1}{2}
\sum_{i=1}^{N}r_i^2+\sum_{i<j}V_{ij},
\eqno{(11)}
$$
where
$$
V_{ij}=
\frac{\gamma_c}{\mid\vec{r}_i-\vec{r}_j\mid}.
\eqno{(12)}
$$
Now we introduce the Jacobi coordinates $ \vec{\zeta}_1=\vec{R}=(1/N)
\sum_{i=1}^{N}\vec{r}_i$, the center-of-mass coordinate, and ($i\ge2$)
$$
\vec{\zeta}_i=\frac{1}{\sqrt{i(i-1)}}[(1-i)\vec{r}_i+\sum_{k=1}^{i-1}\vec{r}_k].
\eqno{(13)}
$$
Using
$$
\sum_{i=1}^{N}r_i^2=NR^2+\sum_{i=2}^{N}\zeta_i^2
\eqno{(14)}
$$
we can rewrite Eq.(11) as
$$
H=-\frac{1}{2N} \Delta_R-\frac{1}{2}\sum_{i=2}^{N} \Delta_{\zeta_i}+
\frac{1}{2}NR^2+\frac{1}{2}\sum_{i=2}^{N}\zeta_i^2
+\sum_{i<j}V_{ij}.
\eqno{(15)}
$$
Hence we have for the ground-state energy
$$
E=\frac{3}{2}+<\psi\mid[-\frac{1}{2}\sum_{i=2}^{N} \Delta_{\zeta_i}+
+\frac{1}{2}\sum_{i=2}^{N}\zeta_i^2+\sum_{i<j}V_{ij}]\mid \psi>,
\eqno{(16)}
$$
where $\psi(\vec{r}_1, \vec{r}_2, ... \vec{r}_N)$ is the ground state-wave 
function.
 Using symmetric properties of $\psi$ we can rewrite Eq. (16) as
$$
E=\frac{3}{2}+<\psi\mid (N-1)(-\frac{1}{2} \Delta_{\zeta_2}+
\frac{1}{2}\zeta_2^2+\frac{N}{2}V_{12}(\sqrt{2}\zeta_2))\mid \psi>.
\eqno{(17)}
$$
Projecting $\mid\psi>$ on the complete basis $\mid n>$,
generated by the effective two-body eigenvalue problem
$$
H^{(0)}\mid n>=(N-1)(-\frac{1}{2} \Delta_{\zeta_2}+
\frac{1}{2}\zeta_2^2+\frac{N}{2}V_{12}(\sqrt{2}\zeta_2))\mid n>=\epsilon_n \mid
n>
,
\eqno{(18)}
$$
we get
$$
E=\frac{3}{2}+\sum_n \epsilon_n \mid<\psi\mid n>\mid^2 \ge(\frac{3}{2}+\epsilon
_0).
\eqno{(19)}
$$
Hence
 the ground state energy of the effective two-body hamiltonian $H^{(0)}$,
$\epsilon_0$, is a lower bound of $E-\frac{3}{2}$.
Eq.(19) is a generalization of the Post and Hall lower-bound method [15]
for the case of system of interacting particles confined in a harmonic trap.
In the particular case of bosons with the Hooke interaction,  this procedure, Eq
.(19),
gives the
exact value of the ground-state energy (see Appendix for details).

To find $\epsilon_0$ for the Coulomb interaction case, Eq.(6), we need to solve
the effective two-body problem
$$
\tilde{H}\phi=-\frac{1}{2} \frac{d^2\phi}{d\zeta^2}+\frac{1}{2}\zeta^2\phi+
\frac{\lambda}{\zeta}\phi=\tilde{\epsilon}\phi,
\eqno{(20)}
$$
where
$\lambda=N \gamma_c/(2 \sqrt{2}),$ and $ \tilde{\epsilon}=\epsilon_0/(N-1)$.

For the case of $\lambda<1$, the weak coupling pertubation (WCP) calculation
leads to the
ground state energy $\tilde{\epsilon}$ given by [24]
$$
\tilde{\epsilon}=\frac{3}{2}+1.128379 \lambda- 0.15578 \lambda^2+...
\eqno{(21)}
$$

\vspace{18pt}

{\bf IV. Strong Coupling Pertubative Expansion}
\vspace{8pt}

The two-body problem with the so-called spiked harmonic oscillator (SHO)
$V(r)=r^2+\frac{l(l+1)}{r^2}+\frac{\lambda}{r^{\alpha}}$, where $r\ge0$,
and $\alpha$ is positive constant, has been the subject of intensive study
[21-28]. The quantity $\lambda$ is a positive definite parameter, it measures
the strength of the pertubative potential. It was found [22] that the normal
pertubation theory could not be applied for the values $\alpha\ge 5/2$,
so-called singular spiked harmonic oscillator. In Ref.[21],  a
special pertubative theory was developed for this case. A  strong coupling pertu
bative
expansion (SCP) ($\lambda>1$)was carried out in Ref.[24].  In Ref.[27] the SCP
was used for the case of $\alpha=3$. In Refs.[23, 26], it
was shown that the SHO problem with $\alpha=1$ is solvable analytically for a
particular set of oscillator frequencies.
For example, for $\lambda=1$,  we have [23]
$$
\tilde{\epsilon}=\frac{5}{2},~~\phi(\zeta)=\zeta e^{-\zeta^2/2}(1+\zeta),
\eqno{(22)}
$$
and for $\lambda=\sqrt{5}$ we have [26]
$$
\tilde{\epsilon}=\frac{7}{2},~~\phi(\zeta)=\zeta e^{-\zeta^2/2}
(1+\sqrt{5} \zeta+\zeta^2).
\eqno{(23)}
$$

Eq.(20) can be solved for the case of large $\lambda$ using the SCP [24].
The idea of this method is to expand the potential
$V(\zeta)=\frac{\zeta^2}{2}+\frac{\lambda}{\zeta}$ around its minimum
$$
V(\zeta)=\frac{3}{2}\lambda^{2/3}+\frac{3}{2}(\zeta-\lambda^{1/3})^2
+\sum_{i=1}^{\infty}(-1)^{i}\frac{\lambda^{-i/3}}{i+2}(\zeta-\lambda^{1/3})^{i+2
}.
\eqno{(24)}
$$
Substitution of Eq.(24) into Eq.(20) gives
$$
\tilde{H}=H_0+H^{\prime},
\eqno{(25)}
$$
where the nonpertubative Hamiltonian $H_0$ is given by
$$
H_0=-\frac{1}{2}\frac{d^2}{dz^2}+\frac{3}{2}\lambda^{2/3}+
\frac{3}{2}z^2,
\eqno{(26)}
$$
and pertubation $H^{\prime}$ is given by
$$
H^{\prime}=\sum_{i}^{\infty}H_i\lambda^{-i/3},
\eqno{(27)}
$$
with $H_i=(-1)^i z^{i+2}/(i+2),$ and $z=(\zeta-\lambda^{1/3})$.

Now $\phi$ and  $\tilde{\epsilon}$ can be written as
 $$\phi=\lim_{n \rightarrow \infty}\phi_n
\eqno{(28)}
$$
and
$$
\tilde{\epsilon}=\lim_{n \rightarrow \infty}\tilde{\epsilon}_n,
\eqno{(29)}
$$
where

$\phi_n=\sum_{i=0}^n \phi^{(i)} \lambda^{-i/3},$
and $\tilde{\epsilon}_n=\sum_{i=0}^n\tilde{\epsilon}^{(i)} \lambda^{-i/3}$.
Substitution Eqs.(26), (28), and (29) into Eq.(20)
gives
$$
\sum_{i=0}^nH_i\phi^{(n-i)} =\sum_{i=0}^n\epsilon^{(i)}\phi^{(n-i)}
\eqno{(30)}
$$
The complete oscillator basis $\mid\tilde{n}>$,
$H_0\mid\tilde{n}>=e_n \mid\tilde{n}>$, where $z=(\zeta-\lambda^{1/3})$
is extended to the full real axis, is used to solve Eq.(30) with $e_0=
\tilde{\epsilon}^{(0)}$, and $\mid 0>=\phi^{(0)}$.
We note that the region $-\infty<z\le-\lambda^{1/3}$ is spurious.
For large $\lambda$, it is expected that the harmonic oscillator basis does
not penetrate too much into forbidden region $z<-\lambda^{1/3}$.
 From Table I, we can see that the SCP converges very fast
for $\lambda>2$.
 However, for the case of $\lambda=1$, it is certainly outside the convergence
radius (see Table II).
Even in this case, $\tilde{\epsilon}_0$ is still  a good lower-approximation
for $\tilde{\epsilon}$.

From the SCP expansion in  the large $\lambda$ limit
we obtain in the large $N$ limit ($N\gg1$, and $N\gamma_c\gg1$)
$$
\epsilon_0=\frac{3}{4}N^{5/3}\gamma_c^{2/3}.
\eqno{(31)}
$$
Combaining Eq.(31) with Eq.(10) we get
in this limit
$$
\frac{3}{4}N^{5/3}\gamma_c^{2/3}\le E\le\frac{9}{10}N^{5/3}\gamma_c^{2/3},
\eqno{(32)}
$$
where $E$ is the leading term of the ground-state energy.
Hence the leading term of the ground state energy in the large N limit is
determined within an accuracy of  $\pm 8\%$.  We can therefore  state that
the mean field theory, Eq.(10), provides a reasonable result in this limit
for the ground-state energy.\\

\vspace{18pt}

{\bf V. Upper Bounds }
\vspace{8pt}

Our method for obtaining the upper bounds, the equivalent linear two-body
equation (ELTBE)  method [16] consists of two
steps. The first
is to give the $N$-body wave function $\psi({\vec{r}_1},{\vec{r}_2}, ...)$
 a particular functional form
$$
\psi(\vec{r}_1, ...\vec{r}_N)\approx \frac{ \Phi( \rho)}{ \rho^{(3N-1)/2}},
\eqno{(33)}
$$
where
$\rho=[\sum_{i=1}^{N}r_i^2]^{1/2}.$

 The second
step is to derive
an equation for $\Phi( \rho)$ by requiring that $\psi({\vec{r}_1},{\vec{r}_2},..
.)$ must satisfy a variational principle
$\delta  <\psi\mid H \mid \psi> = 0,
$
with a subsidiary condition $<\psi\mid \psi>=1$.$~~$$H$ is the
Hamiltonian.
This leads to the following equation
$$
H_{\rho}\Phi=[-\frac{1}{2}\frac{d^2}{d \rho^2}+\frac{1}{2}\rho^2+
\frac{(3N-1)(3N-3)}{8 \rho^2}+\frac{\tilde{\lambda}}{\rho}]\Phi=\tilde{E}\Phi,
\eqno{(34)},
$$
where $$\tilde{\lambda}=
     \frac{2}{3 \sqrt{2\pi}} \gamma_c N \frac{\Gamma(3N/2)}{\Gamma(3N/2-3/2)}.
\eqno{(35)}
$$
The lowest eigenvalue of $H_{\rho}$ (Eq.(34)) is an upper bound of the lowest ei
genvalue
of the original $N$-body problem.
Since  a variational estimate of the lowest eigenvalue of $H_{\rho}$ is also
an upper bound of the ground-state energy of the original $N$-body problem,
we have for  this
upper bound, $E_{upper}$ the following expression
$$
E_{upper}=\frac{<\Phi_t\mid H_{\rho}\mid\Phi_t>}{<\Phi_t\mid\Phi_t>}.
\eqno{(36)}
$$
Assuming the following form for the trial function $\Phi_t$,
$$
\Phi_t(\rho)=\rho^{(3N-1)/2} e^{-\rho^p/(2 \alpha^p)},
\eqno{(37)}
$$
we obtain
$$
E_{upper}=\frac{p(3N-2+p)\Gamma((3N-2)/p+1)}{8\Gamma(3N/p)\alpha^2}+
\frac{\Gamma((3N+2)/p)}{2\Gamma(3N/p)}\alpha^2+
\frac{\tilde{\lambda}\Gamma((3N-1)/p)}{\Gamma(3N/p)\alpha},
\eqno{(38)}
$$
where parameters $\alpha$ and $p$ are to be determined from solution of the following equations
$$
\frac{\partial E_{upper}}{\partial \alpha}=\frac{\partial E_{upper}}{\partial p}=0.
\eqno{(39)}
$$

 From Table III, we can see that for the case of $N \gamma_c\le 100$,
the calculated bounds determine the actual value of the ground state energy
within  $\pm \Delta$ accuracy, with $\Delta<9\%$.

\vspace{18pt}

{\bf VI. Large $\gamma_c$ Limit }
\vspace{8pt}

To make a proper choice for the large $\gamma_c$ limit of the Hamiltonian,
 Eq.(11),
we rescale variables, $\vec{r} \rightarrow \gamma_c^{1/3} \vec{r}$, and write 
the
 Schr\"odinger equation for $N$ identical charged bosons confined in a harmonic
 isotropic trap as
$$
[-\frac{1}{2\gamma_c^{(4/3)}}\sum_{i=1}^N \Delta_i+
\frac{1}{2}\sum_{i=1}^N r_i^2+\sum_{i<j} \frac{1}{\mid\vec{r}_i-\vec{r}_j\mid}]
\psi=\frac{E}{\gamma_c^{2/3}}\psi.
\eqno{(40)}
$$
Eq.(40) describes the motion of $N$ particles with an effective mass $\gamma_c^{
4/3}$. Therefore, when $\gamma_c\rightarrow\infty$, the effective mass of the 
particles
 becomes infinitely large and then the particles may be assumed to remain
 essentially stationary at the absolute minimum of the potential energy
$$
V_{eff}(\vec{r}_1,...\vec{r}_N)=
\frac{1}{2}\sum_{i=1}^N r_i^2+\sum_{i<j} \frac{1}{\mid\vec{r}_i-\vec{r}_j\mid},
\eqno{(41)}
$$
with quantum fluctuations around the classical minimum. Obviously this 
assumption
 fails if the potential energy $V_{eff}$ does not possess a minimum and (or) 
gradients of the wave functions are large. This large $\gamma_c$ limit is the
Wigner cristallization regime [6].

Interest in the investigation of the Wigner crystallized ground state has grown as a result of recently proposed quantum computer by Cirac and Zoller [29].
(See also Refs. [30-33]).

As we have already noted in Sec. II, mean-field theory, Eq.(1), can not describe crystallized ground state. Therefore we  only  can state that mean-field ground-state energy is an upper bound to the exact energy. Strightforward calculations for the case of $\gamma_c\gg1$ give the Thomas-Fermi upper bound
$$
E_{upper}=\frac{9}{10}N(\gamma_c(N-1))^{2/3}.
\eqno{(42)}
$$
From the SCP expansion, Eq.(24), we obtain in the large $\gamma_c$ limit a lower bound
$$
E_{low}=\epsilon_0=\frac{3}{4}(N-1)(N\gamma_c)^{2/3}.
\eqno{(43)}
$$
Therefore for the leading term of the ground-state energy, $E$, we have
$$
\frac{3}{4}(N-1)(N \gamma_c)^{2/3}\leq E\leq\frac{9}{10}N(\gamma_c(N-1))^{2/3}.
\eqno{(44)}
$$
From Eq.(44) we can see that in the case of the Wigner crystallization regime, $\gamma_c\gg1$, our bounds determine the ground-state energy within $\pm\Delta$
accuracy, with $\Delta\approx8\%$ for $N \geq 100$, $\Delta\approx10\%$
for $N=10$ and $\Delta\approx15\%$ for$N=3$.
It shows that the mean-field theory, Eq.(10) provides a reasonable upper bound for $N>10$ even in the large $\gamma_c$ limit. However the Thomas-Fermi treatment can not describe the crystallized ground-state wave function, since a small relative error of the mean-field ground-state energy does not necessarily imply that the mean-field (product) state describes the actual many body wave function well.
\vspace{18pt}

{\bf VI. Summary and Conclusion}
\vspace{8pt}

In summary, we have generalized the Post and Hall lower-bound method [15] for
the case of interacting bosons confined in a harmonic trap.

As examples of application, we have studied bosons interacting with Coulomb
forces in a harmonic trapping potential. We have found the upper bounds  using
the mean-field approach and the ELTBE method [16].

It is shown that the leading term of the ground state energy in the large $N$ 
limit
 ($N\gg 1$ and $N\gamma_c\gg1$) is determined within an accuracy of
$\pm 8\%$, and it is also  shown that the mean-field theory provides a 
reasonable
 results with relative error of less than 16\% for the leading term of ground 
state energy.

However the Thomas-Fermi treatment can not describe the crystallized ground-state wave function, since a small relative error of the mean-field ground-state energy does not necessarily imply that the mean-field (product) state describes the actual many-body wave function well.
\vspace{18pt}

{\bf Appendix}
\vspace{8pt}

In this Appendix we consider Hamiltonian [34-35]
$$
H=-\frac{1}{2}\sum_{i=1}^{N} \Delta_{i}+\frac{1}{2}
\sum_{i=1}^{N}r_i^2+\frac{\Lambda}{2}\sum_{i<j}(\vec{r}_i-\vec{r}_j)^2,
\eqno{(A.1)}
$$
which was used for a problem in nuclear physics in Ref. [36].

Using Eq.(14) and
$$
\sum_{i<j}(\vec{r}_i-\vec{r}_j)^2=N \sum_{i=2}^{N}\zeta_i^2,
\eqno{(A.2)}
$$
we can rewrite Eq.(A.1) as
$$
H=-\frac{1}{2N} \Delta_R+\frac{1}{2}NR^2
+\sum_{i=2}^{N}[-\frac{1}{2}\Delta_{\zeta_i}+\frac{1+N \Lambda}{2}\zeta_i^2]
\eqno{(A.3)}
$$
This leads to the ground-state energy
$$
E=\frac{3}{2}[1+\sqrt{1+N \Lambda}(N-1)],
\eqno{(A.4)}
$$
which is equal to the lower bound, Eq.(19), with
$$
\epsilon_0=\frac{3}{2}\sqrt{1+N \Lambda}(N-1).
\eqno{(A.5)}
$$

\pagebreak

TABLE I. Results for ground-state energy, $\tilde{\epsilon}$ (Eq.20).
We compare zero order, econd order and converged results (10th order) to the
exact analytical solution (Eqs.(22-23)).
\vspace{8pt}

\begin{tabular}{lllll}
\hline\hline
$\lambda$
&$\tilde{\epsilon}_0$
&$\tilde{\epsilon}_2$
&$\tilde{\epsilon}_{converged}$
&$\tilde{\epsilon}_{exact}$ \\ \hline
1
&$~~$2.36603
&$~~$2.46325
&
&2.5 \\ \hline
$\sqrt{5}$
&$~~$3.43099
&$~~$3.48785
&$~~$3.49954
&3.5 \\ \hline
10
&$~~$7.82841
&$~~$7.84935
&$~~$7.85061
& \\ \hline
100
&$~$33.18255
&$~$33.18705
&$~$33.18711
& \\ \hline
500
&$~$95.3601
&$~$95.36165
&$~$95.36165
& \\ \hline
1000
&150.86603
&150.86700
&150.86700
& \\ \hline
5000
&439.46869
&439.46902
&439.46902
& \\ \hline
10000
&697.10435
&697.10456
&697.10456
&\\  \hline\hline
\end{tabular}\\
\vspace{18pt}

TABLE II. Results for $\tilde{\epsilon}_n$ for the $\lambda=1$ case.
\vspace{8pt}

\begin{tabular}{lllllll}
\hline\hline
$\lambda$
&$\tilde{\epsilon}_0$
&$\tilde{\epsilon}_2$
&$\tilde{\epsilon}_4$
&$\tilde{\epsilon}_6$
&$\tilde{\epsilon}_8$
&$\tilde{\epsilon}_{10}$\\ \hline
1
&2.36603
&2.46325
&2.48797
&2.49716
&2.50439
&2.5125 \\  \hline\hline
\end{tabular}\\
\vspace{18pt}

\pagebreak

TABLE III. Results for upper, $E_{upper}/N$, lower, $E_{lower}/N$
bounds of ground state energy per particle, and $\Delta=
(E_{upper}-E_{lower})/(2 E_{upper})$.
\vspace{8pt}

\begin{tabular}{lllll}
\hline\hline
$N$
&$\lambda=N\gamma_c/(2\sqrt{2})$
&$E_{lower}/N$
&$E_{upper}/N$
&$\Delta, \%$\\ \hline
10
&0.1
&$~$1.60015
&$~$1.60048
&0.02 \\ \cline{2-5}

&0.5
&$~$1.97272
&$~$1.98724
&0.4 \\ \cline{2-5}

&1
&$~$2.4
&$~$2.43945
&0.8 \\ \cline{2-5}

&$\sqrt{5}$
&$~$3.3
&$~$3.4478
&2.1 \\ \cline{2-5}

&10
&$~$7.21555
&$~$8.18751
&5.9 \\ \cline{2-5}

&100
&30.0184
&36.8931
&9.3 \\ \hline
100
&0.1
&$~$1.61017
&$~$1.61068
&0.02 \\ \cline{2-5}

&0.5
&$~$2.01999
&$~$2.03468
&0.36 \\ \cline{2-5}

&1
&$~$2.49
&$~$2.52904
&0.8 \\ \cline{2-5}

&$\sqrt{5}$
&$~$3.48
&$~$3.62737
&2.0 \\ \cline{2-5}

&10
&$~$7.7871
&$~$8.76512
&5.6 \\ \cline{2-5}

&100
&32.8702
&39.8116
&8.7  \\ \hline\hline
\end{tabular}\\

\pagebreak

\begin{center}
{\bf References}
\end{center}

\vspace{8pt}

\noindent
[1] M. F. M. Osborne, Phys. Rev. {\bf 76}, 400 (1949).

\noindent
[2] M. R. Schafroth, Phys. Rev. {\bf 100}, 463 (1955).

\noindent
[3] L. L. Foldy, Phys. Rev. {\bf 124}, 649 (1961).

\noindent
[4] K. A. Brueckner, Phys. Rev. {\bf 156}, 204 (1967).

\noindent
[5] A. S. Alexandrov, W. H. Beere, and V. V. Kabanov, Phys. Rev. B{\bf 54},
 15363 (1996).

\noindent
[6] T. Schneider and R. Bl\"umel, J. Phys. B{\bf 32}, 5017 (1999).

\noindent
[7] Y.E. Kim, and A.L. Zubarev, Eur. Phys. Journal D (submitted).

\noindent
[8] B. D. Esry, Phys. Rev. A{\bf 55}, 1147 (1997).

\noindent
[9] J. J. Thomson, Philos. Mag. {\bf 7} (Sixth Series), 237 (1904).

\noindent
[10] Yu. E. Lozovik, Usp. Fiz. Nauk. {\bf 153}, 356 (1987) [Sov. Phys. Usp.
{\bf 30}, 912 (1987)].

\noindent
[11]  F. Diedrich, E. Peik, J. M. Chen, W. Quint, and  H. Walter,
Phys. Rev. Lett. {\bf 59}, 2931 (1987); D. J. Wineland, J. C. Bergquist,
W. M. Itano, J. J. Bolinger, and C. H. Manney, Phys. Rev. Lett. {\bf 59}, 2935
(1987).

\noindent
[12] M. A. Read and W. P. Kirk, {\it Nanostructure Physics and Fabrication}
(Academic Press, Boston, 1989);
L. Jacak, P. Hawrylak and A. Wojs, {\it Quantum Dots} (Springer, Berlin, 1998);
T. Chakraborty, {\it Quantum Dots} (Elsevier, 1999);
D. Bimberg, M. Grundmann and N.N. Ledentsov, {\it Quantum Dot Heterostructures} (John Wiley \& Sons, 1999).

\noindent
[13] M. A. Kastner, Phys. Today {\bf 46}, 24  (1993).

\noindent
[14] E. H. Lieb, R. Seiringer and J. Yngvason, Phys. Rev. A{\bf 61}, 043602
(2000).

\noindent
[15] H. R. Post, Proc. Phys. Soc. A{\bf 69}, 936 (1956);
 H. R. Post, Proc. Phys. Soc. {\bf 79}, 819 (1961);
R.L. Hall, and H.R. Post, Proc. Phys. Soc. {\bf 90}, 381 (1967).

\noindent
[16] A. L. Zubarev and Y. E. Kim, Phys. Lett. A{\bf 263}, 33 (1999);Y. E.
Kim  and A. L. Zubarev, J. Phys. B: At. Mol. Opt. Phys. {\bf 33}, 55 (2000).

\noindent
[17] E.P. Wigner, Phys. Rev. {\bf 46}, 1002 (1934).

\noindent
[18] D. A. Huse and E. D. Siggia, J. Low. Temp. Phys. {\bf 46}, 137 (1982).

\noindent
[19] G. Baym and C. J. Pethick, Phys. Rev. Lett. {\bf 76}, 6 (1996).

\noindent
[20] E. Lieb, Rev. Mod. Phys. {\bf 53}, 603 (1981).

\noindent
[21] E.M. Harrell, Ann. Phys. (NY) {\bf 105},379 (1977).

\noindent
[22] L.C. Detwiler and J.R. Klauder, Phys. Rev. D{\bf 11}, 1436 (1975);
J.R. Klauder, Science, {\bf 199}, 735 (1978).

\noindent
[23] S. Kais, D.R. Hershbach, and R.D. Levine, J. Chem. Phys.
{\bf 91}, 7791 (1989).

\noindent
[24] V.C. Aguilera-Navarro, G.A. Esterez, and R. Guardiola, J. Math. Phys.
{\bf 31}, 99 (1990); V.C. Aguilera-Navarro and  R. Guardiola, J. Math. Phys.
{\bf 32}, 2135 (1991).

\noindent
[25] M. Znoil, Phys. Lett. A {\bf 158}, 436 (1991).

\noindent
[26] M. Taut, Phys. Rev. A{\bf 48}, 3561 (1993).

\noindent
[27] D.K. Watson and B.A. McKinney, Phys. Rev. A{\bf 59}, 4091 (1999).

\noindent
[28] R.L. Hall and N. Saad, J. Phys. A: Math. Gen. {\bf 33}, 5531 (2000).

\noindent
[29] J.I. Cirac and P. Zoller, Phys. Rev. Lett. {\bf74}, 4094 (1995).

\noindent
[30] A. Steane, Appl. Phys. B{\bf 64} 623 (1977).

\noindent
[31] Special Issue on Quantum Information [Phys. World {\bf 11}, No. 3, 33-57 (1998)].

\noindent
[32] D.F.V. James, Appl. Phys. B{\bf 66}, 181 (1998).

\noindent
[33] A. S\o rensen and K. M\o lmer, Phys. Rev. Lett.  {\bf 82}, 1971 (1999).

\noindent
[34] H.R. Post, Proc. Phys. Soc. A{\bf 66}, 649 (1953).

\noindent
[35] N.F. Johnson and M.C. Payne, Phys. Rev. Lett. {\bf67}, 1157 (1991);
A.L. Zubarev and V.B. Mandelzweig, Phys. Rev. C{\bf 50}, 38 (1994);
F. Brosens, J.T. Devreese and L.F. Lemmens, Phys. Rev. E{\bf55}, 6795 (1997);
N.K. Wilkin, J.M. Gunn and R.A. Smith, Phys. Rev. Lett. {\bf80}, 2265 (1998).

\noindent
[36] S. Gartenhaus and C Schwartz, Phys. Rev. {\bf108}, 482 (1957).
\end{document}